# AN ADAPTIVE NETWORK-BASED APPROACH FOR ADVANCED FORECASTING OF CRYPTOCURRENCY VALUES

Ali Mehrban[1] and Pegah Ahadian[2]

[1]School of Electrical and Electronic Engineering, Newcastle University, Newcastle, UK
a.mehrban@ieee.org
[2]Department of Computer Science, Kent State University, Ohio, USA
pahadian@kent.edu

## ABSTRACT

*This paper describes an architecture for predicting the price of cryptocurrencies for the next seven days using the Adaptive Network Based Fuzzy Inference System (ANFIS). Historical data of cryptocurrencies and indexes that are considered are Bitcoin (BTC), Ethereum (ETH), Bitcoin Dominance (BTC.D), and Ethereum Dominance (ETH.D) in a daily timeframe. The methods used to teach the data are hybrid and backpropagation algorithms, as well as grid partition, subtractive clustering, and Fuzzy C-means clustering (FCM) algorithms, which are used in data clustering. The architectural performance designed in this paper has been compared with different inputs and neural network models in terms of statistical evaluation criteria. Finally, the proposed method can predict the price of digital currencies in a short time.*

## 1. INTRODUCTION

Digital currency is a form of electronic money that operates on the internet and possesses most of the attributes of conventional money, except for its physical absence. A subset of digital currency is cryptocurrency, which is encrypted by specific algorithms. These cryptocurrencies often utilize blockchain technology to record transactions [1]. The main distinction between cryptocurrencies and other digital currencies is the level of security of the former. Some examples of digital currencies are Web Money and Perfect Money, and some examples of cryptocurrencies are BTC and ETH.

A trade or transaction is one of the main economic concepts that includes the purchase and sale of goods and services by paying a fee from the buyer to the seller or the exchange of goods or services between the parties to the transaction. The network that facilitates trading is called the market. In financial markets, trading refers to the buying and selling of securities, such as buying stocks from a stock exchange or buying and selling cryptocurrencies. In trading, the general goal is to make a profit by buying at a lower price and selling at a higher price, usually in a short period.

Technical analysis (TA) is a key technique for financial market analysis, enabling optimal transaction timing. However, TA proficiency alone is insufficient for market success; time commitment and market vigilance are also required. Additionally, human decision-making differs from algorithmic processes, as it is influenced by biological and psychological factors (e.g., fatigue, fear, greed), which may impair its effectiveness.

Cryptocurrencies are subject to severe price fluctuations in the financial market because they are independent of any governmental organization. These fluctuations create opportunities for large profits but also expose traders to high risks [1]. Hence, automated systems are required in the financial market. The existing automated systems for the Forex market are inadequate for this market, so a market-specific automated system is necessary. According to a research, automated trading systems dominate most of the current trades [2].

These systems enable traders to define rules for capital entry and exit that can be automatically executed by computer. These systems are order execution methods that use automated trading instructions and pre-programmed algorithms to compute variables such as price, time, and volume. Algorithmic trading to perform the specified strategy correctly and completely perform tasks that are sometimes performed entirely automatically by robots (trading robots) that include "fully automated" transactions [4], [5] and sometimes in some areas of taste. And human opinion is involved, in which case transactions are referred to as "semi-automated" [6], [7]. semi-automated algorithms are classified into several categories: Trading Algorithms, Signaling Algorithms, Monitoring Algorithms, Position Trading Algorithms, and Frequency Algorithms. In contrast, the disadvantages of these systems can be low coding accuracy, mechanical defects, monitoring, and over-optimization.

Signaling algorithms enhance market analysis and decision-making, boosting traders' profits. They can signal price or trend forecasts. Price forecasts are short-term profitable [8], but trend forecasts are long-term effective. Price forecasting systems use machine learning on time-series data [9], while trend forecasting systems use technical analysis indices. Price forecasting can improve trading outcomes, but price prediction is challenging [2].

## 2. RELATED WORK

Signaling algorithms provide the analyst with more information about market conditions and help the analyst to improve the flow of analysis and decision-making and consequently trades, which results in increased profitability of the trader or analyst. These algorithms can signal both price prediction and price trend prediction.

Price prediction is one of the most important issues in financial and commercial markets, we can decide to buy or sell knowing the future price, so extensive research has been done in this area. Price data in financial markets are of the type of time series data, time series is a set of data that occurred in a period, respectively. Machine learning techniques can be used to predict time series data. Time series forecasting methods use historical and statistical information to forecast, so these methods can also be used to predict prices.

Because price prediction is a difficult nonlinear problem, neural networks (NN) can be used to solve this problem. Reference [10] uses an artificial neural network (ANN) to predict stock prices in the Hong Kong Stock Exchange. The innovation of this article is in data preprocessing, the steps of this preprocessing are in the form of clearing incomplete data, and normalizing the data using Max-Min and Z-Score methods, Finally, the results indicate that the normalization of data using the Max-Min method in the interval $\left[-\frac{1}{2}, \frac{1}{2}\right]$ compared to other intervals and the Z method -Score lowers the Root Mean Square Error (RMSE) benchmark. A fully automated network trading system as a pipeline is designed for the Forex market in [5]. In this pipeline, a regression network with Scaled Conjugate Gradient (SCG) algorithm is used to predict the price in the initial block, then the forecasted and is given to the next blocks to determine the price trend and decide. In this report, the system has improved the Maximum Drawdown (MDD) and Return on Investment (ROI) benchmark compared to a system that used the Ichimoku Index with a resistance and support line to find the right position for entry and exit.

As mentioned, machine learning techniques can predict time series and price. In [6] BTC information is collected in daily timeframes and then with machine learning algorithms such as

Linear Regression (LR), Logical Regression (LGR), Multivariate Regression (MR), Multiple Linear Regression (MLR), Least Squares Partial (PLS), Postpartum Error Neural Network (BPNN), KNN and DT and SVM and Regression Support Vector (SVR), Gradient Boosting Machine (GBM), Lasso Regression (LASSO), price is forecast for BTC, finally price predicted for BTC by LASSO algorithm has the best performance in terms of RMSE and MAE criteria. In [7] BTC, ETH, and Ripple trading data are preprocessed to remove noise data and then normalized for easier processing in the range of [0, 1] The data were then analyzed using several data mining methods including Decision Tree (DT) and K Nearest Neighbor (KNN) and Support Vector Machine (SVM) and NN, The results show that the SVM had the best and the DT the worst performance in price forecasting in terms of MAE and RMSE criteria.

BTC, Ripple, and ETH trading data are collected in [1]. A comparison of this collection shows that the price increase of BTC has been positive on the price increase of Ripple and ETH, then it uses two nonlinear algorithms, DT and KNN to predict the price of cryptocurrencies. The results of the experiments show that the k-closest neighborhood model performs better in terms of MAE and MSE criteria in the low set space than the predictive DT model.

An Adaptive Neuro-Fuzzy Inference System or Adaptive Network-Based Fuzzy Inference System (ANFIS) is a kind of artificial neural network that is based on the Takagi–Sugeno fuzzy inference system. it integrates both neural networks and fuzzy logic principles, its inference system corresponds to a set of fuzzy IF-THEN rules that have the learning capability to approximate nonlinear functions. Because price prediction is a difficult nonlinear problem, we can use it to price prediction. For prediction using ANFIS, In [11] the data are given to a group of compatible ANFIS and then the output of each of these systems is combined using the average and weighted averages.

The main idea of this method is that each ANFIS system uses a different method for training and membership performance. In this method, the Mackey-Glass time series is predicted with 98% accuracy. It is designed in [12] systems for forecasting stock market prices called PATSOS. This system uses two consecutive ANFIS systems, the first of which is called ANFIS Controller (CON-ANFIS) and the second is ANFIS Process (PR-ANFIS) and it has feedback to the controller. Finally, according to the results, this system has a higher value in terms of ROR than the buy and hold (B&H) strategy. Reference [13] uses the system designed in [12] to predict the price of cryptocurrencies using various types of membership functions Finally, the results show that the RMSE criterion for PATSOS system is lower than ANFIS and ANN.

The deep learning method refers to powerful machine learning algorithms that specialize in solving nonlinear and complex problems by utilizing large amounts of data to become efficient predictive models. Since price prediction is a nonlinear problem, a deep learning method can be used to solve this problem. In [2], BTC, ETH, and Ripple cryptocurrencies were first obtained from the Kraken cryptocurrency market and then using a combination of three models of deep learning, Long Short-Term Memory (LSTM) and Bidirectional Long-Short-Term Memory Networks (BiLSTM) and Convolution Neural Networks (CNN) offers its own advanced deep learning models called CNN-LSTM and CNN-BiLSTM, then predicts the price using the said deep learning models and designed models and machine learning models SVR, KNN and DTR and compares their results by the evaluation criteria of MAE, RMSE, ACC, and F1-Score.

In summary, the advanced deep learning models perform slightly better than the machine learning models, while the combined deep learning models perform better than the other two models, but this performance is not significantly different and there is not much change in performance, So the idea arises that it is likely that the cryptocurrency price is a random walk or very close to it, so deep learning does not help much in learning price patterns.

# 3. METHODOLOGY

As mentioned, it is impossible to avoid financial markets today with the advancement of technology, finding suitable entry and exit points will increase profits and reduce trading losses in these markets, so the decision here is more important. It was also said that today most of these are automated systems that are trading in the financial markets. In addition to the payment method, cryptocurrencies are a good way to invest, but due to their nature, systems designed in other markets are not applicable to them, so designing an automated system suitable for this market is important. As we have seen, because price data are a type of time series data, they can be predicted using machine learning methods.in this paper, we present a system for predicting the price of cryptocurrencies using the ANFIS system. The block diagram of this system is shown in Fig. 1.

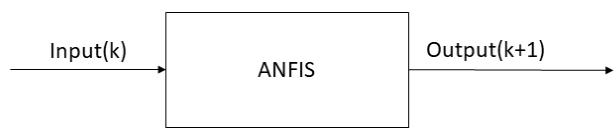

Fig. 1.    ANFIS block diagram.

## A. *Adaptive Neuro-Fuzzy Inference System*

Adaptive Neuro-Fuzzy Inference System or Adaptive Network-Based Fuzzy Inference System is an adaptive fuzzy inference system that is a feedforward neural network with training capabilities [11]. This system combines neural networks with the principles of fuzzy logic, so it has the advantages of both, its inference system corresponds to a set of ambiguous IF-THEN rules that have the ability to learn the approximation of nonlinear functions [12], [14].

### 1) Structure of ANFIS

The ANFIS structure consists of 5 layers. In the following, we describe the structure of the Sugeno-type ANFIS system with two inputs *x*, *y,* and one output *f* [14]. If the rules are as follows:

$$Rule: if\ x\ is\ A_1\ and\ y\ is\ B_1\ then\ f_1 = p_1 x + q_1 y + r_1 \quad (1)$$

$$Rule: if\ x\ is\ A_2\ and\ y\ is\ B_2\ then\ f_2 = p_2 x + q_2 y + r_2 \quad (2)$$

The layers of the ANFIS system are as follows:

*Layer 1:* The first layer of an ANFIS network describes the difference from vanilla NN. The first layer of which receives input data and at the end determines the membership functions, which is called "Fuzzification". The membership degrees of each function are computed by using the premise parameter set, namely $\{a, b, c\}$.

$$O_{1,i} = \mu A_i(x) \qquad for\ i = 1,2 \qquad (3)$$

$$O_{1,i} = \mu B_{i-2}(x) \qquad for\ i = 3,4 \qquad (4)$$

***Layer 2:*** The output of the second layer is the multiplication of its inputs, which shows the power of each rule. It is equivalent to the "if" part of the rules, which is called the "rule layer".

$$O_{2,i} = w_i = \mu A_i(x)\mu B_i(x) \qquad i = 1,2 \qquad (5)$$

***Layer 3:*** This layer normalizes the output of the second layer, which is called the "normalization (N) layer".

$$O_{3,i} = \overline{w}_i = \frac{w_i}{w_1 + w_2} \qquad i = 1,2 \qquad (6)$$

***Layer 4:*** This layer defuzzifying the values given by the third layer. which is called "Defuzzification". and $\{p_i, q_i, r_i\}$ are a set of consequent parameters.

$$O_{4,i} = \overline{w}_i f_i = \overline{w}_i(p_i x + q_i y + r_i) \qquad i = 1,2 \qquad (7)$$

***Layer 5:*** This layer sums the input signals and produces the final output of the ANFIS system.

$$O_{5,i} = \sum_i \overline{w}_i f_i = \frac{\sum_i w_i f_i}{\sum_i w_i} \qquad i = 1,2 \qquad (8)$$

*2) Data clustering in ANFIS*

In ANFIS, data can be clustered in various ways such as grid partition, subtractive clustering, and FCM. FCM is a type of clustering in which each point does not necessarily belong to just one cluster while the points in this method, using the membership score between [0,1] fuzzy belong to a different cluster.

*3) ANFIS system training methods*

ANFIS modifies and adjusts learnable parameters to minimize the error between the system's actual output and the desired output. One method, the Hybrid algorithm, uses two passes to reduce error. The forward pass sets the consequent parameters using LSE, and the backward pass changes the premise parameters using GD. Another method for tuning the ANFIS system parameters is the Back-propagation algorithm, which uses GD to adjust all parameters. [15].

*B. Dominance*

Dominance in the term means domination. The dominance index is a general index that shows your unique share based on the position and weight of the asset relative to the entire cryptocurrency market. The BTC Dominance (BTC.D) Index is provided by CoinMarketCap[1] and measures the total market value of BTC relative to the total market value of cryptocurrency assets. The market value of BTC is an important index of investor sentiment. Therefore, using this index to predict the price of cryptocurrencies can be appropriate.

*C. The proposed model*

While predictive analytics such as employing SVD within convolutional neural networks (CNNs) have a stronghold in modeling, particularly evident in medical science [17], [18], this paper takes a unique stance by forecasting cryptocurrency values using the adaptive network-based fuzzy inference system (ANFIS). It introduces a novel approach to predictive analysis within this specific domain. The convergence of advanced language models in data analysis echoes the complexities encountered in forecasting cryptocurrency values. This integration highlights the essential depth of algorithms across diverse data domains, emphasizing the necessity for sophisticated methodologies in handling intricate predictive analytics [19], [20]. This paper aims to forecast the cryptocurrency price for the next week, using the method illustrated in the block diagram in Fig. 2, which consists of several simple subsystems. Each subsystem receives the input value and the actual target value and trains the ANFIS system to predict the output in (k+1) based on the input in (k). Since each cryptocurrency may depend on or be affected by the price trends of other cryptocurrencies or indices such as BTC.D, we also need to forecast this signal simultaneously. Therefore, we include subsystems with multiple inputs in this system. We repeat this process for seven cycles by feeding the predicted price back to the subsystems as input, until we obtain the final output as the price prediction for the seventh day.

## 4. EVALUATION

*A. Data set*

This paper analyzes the historical data of BTC and ETH and BTC.D and ETH.D at the end of a daily candle. This historical data has been collected from August 17, 2017, to June 3, 2022, from Tradingview[2]. Education and validation data are divided into two categories, data can be divided in different ways: 50:50, 60:40, 70:30, 80:20, and 90:10. In this study, we consider 90% of the data for training and 10% for testing. Based on data sharing, the models are trained using a training set, and their ability to predict with the test set is evaluated.

*B. Check the parameters*

In this paper, our goal is to use the ANFIS system to forecast the price of BTC and ETH for the following week. We first predict the price of BTC and ETH in the daily time frame at (k+1) by using various parameters, training and clustering methods, and dependencies such as BTC.D and ETH.D. Then, based on Table 1, we keep some parameters of the ANFIS system constant and vary others, such as the clustering method and the training method, to find the optimal method.

---

[1] CoinMarketCap is the world's most trusted and accurate source for cryptocurrency market capitalizations, pricing, and information. CoinMarketCap is a U.S. company registered in the United States of America.

[2] Tradingview is an online trading and charting platform that also acts as a social network. It is an advanced financial visualization platform

As we can see in Table 2, in predicting the price of BTC, when we use only the price of BTC itself to predict its price, according to the RMSE and Root Mean Square Relative Error (RMSRE) criteria, the Backpropagation and FCM methods have the best performance in test data. As we said, the value of BTC.D can affect the emotions of traders and the price of BTC, so once again we get the price of BTC using two inputs, the price of BTC and the amount of BTC.D. As we can see in Table 2, in general, the use of BTC.D has a positive effect on predicting the price of BTC and has improved the RMSE and RMSRE evaluation benchmarks. In Fig. 3 (a), we can see the difference between the actual value and the predicted amount of BTC in the case where we use BTC and BTC.D for prediction, in the testing data, and in Fig. 3 (b), We can see the amount of prediction error, that is, the difference between the actual price and the price predicted by the ANFIS system.

Our aim is to forecast the price of ETH. We first use its price to predict ETH, and then we also include ETH.D and apply the parameters of Table 1 and the optimal methods of Table 2 for prediction. Table 3 shows that the evaluation criteria improve when we add ETH.D to the price of ETH for forecasting. We contrast the ANN method with the ANFIS system in Table 4. In this comparison, our ANN has 10 neurons and uses the Levenberg-Marquardt algorithm for data training, and our ANFIS method uses the fixed parameters of Table 1 and the best-performing parameters in Table 2 for evaluation criteria (FCM method for data clustering and Backpropagation for data training). Both systems use BTC, BTC, and BTC.D as input for predicting BTC, and ETH, ETH, and ETH.D as input for predicting ETH. They also use 90% of the data for training and 10% for testing. The ANFIS system outperforms the ANN method in terms of RMSE criteria for BTC prediction in both training and testing data, but the ANN method does better in ETH prediction in testing data.

### C. Predication for the next seven days

Our aim is to forecast the price of BTC for the following week using the block diagram in Fig. 2. We have seen that BTC.D improves the system evaluation criteria, so we use the block diagram in Fig. 2 with two subsystems, each predicting a signal. The first system predicts the price of BTC based on the two inputs, BTC and BTC.D and the second system predicts the value of BTC.D. We feed the predicted value from each subsystem back to the inputs for the next cycle. We repeat this for seven cycles to forecast the next week. In Fig. 4 (a), we observe the forecasted values for the price of Bitcoin for the next week for test data, with the predicted price diverging from the actual price each day. We also do this process to forecast the price of ETH for the next week. We have seen that ETH.D improves the evaluation criteria, so we use two subsystems for ETH, one predicting ETH.D and the other predicting ETH. In Fig. 4 (b), we see that the forecasted price for ETH differs from the actual price each cycle, with some error.

### D. Statistical performance measures

There are several criteria for evaluating the performance of price forecasting systems. In price forecasting systems, due to the importance of correctly identifying the predicted price, the focus is more on the difference between the actual price and the predicted price and the amount of error. They conclude that we will describe two criteria in the following [6].

*1) Root Mean Square Error:* The normalized distance between the predicted ($z_i$) and actual ($d_i$) values.

$$RMSE = \sqrt{\frac{1}{N}\sum_{i=1}^{N}(d_i - z_i)^2} \qquad (9)$$

*2) Root Mean Square Relative Error:* In fact, each error value obtained from the difference between the actual value and the predicted value is measured relative to the actual value [16].

$$RMSRE = \sqrt{\frac{1}{N}\sum_{i=1}^{N}\left(\frac{d_i - z_i}{z_i}\right)^2} \qquad (10)$$

**TABLE I. Characteristics of ANFIS system**

| Variable | Value/type |
|---|---|
| Type of fuzzy inference system | Sugeno |
| Input | BTC(k) |
| Output | BTC (k+1) |
| Data time frame | daily |
| Membership function type | Gaussmf |
| Logical operation | AND |
| Output membership function type | Linear |
| Number of training data | 1700 |
| Training error goal | 0 |
| Defuzzification | wtaver |

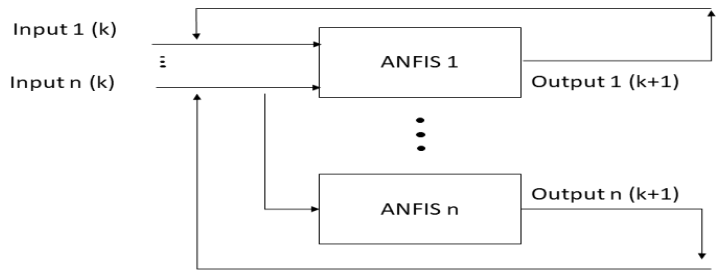

Fig. 2. Price forecasting system for several consecutive times using ANFIS system.

**TABLE II. Bitcoin price forecast result**

| Input | Clustering type | training methods | performance measures | | | |
|---|---|---|---|---|---|---|
| | | | RMSE | | RMSRE | |
| | | | Train | Test | Train | Test |
| BTC(k) | Grid partitioning | Hybrid | **987.28** | 1399.93 | 0.0003802 | 0.002105 |
| | | Backpropagation | 992.02 | 1367.89 | 0.0003788 | 0.002060 |
| | Sub clustering | Hybrid | 992.00 | 1366.12 | **0.0003779** | 0.002062 |
| | | Backpropagation | 993.26 | 1362.18 | 0.0003783 | 0.002056 |
| | FCM clustering | Hybrid | **991.08** | 1364.80 | 0.0003809 | 0.002058 |
| | | Backpropagation | 998.71 | **1335.40** | 0.0003786 | **0.002015** |
| BTC(k), BTC.D(k+1) | Grid partitioning | Hybrid | **928.22** | 2518.68 | **0.0003555** | 0.003941 |
| | | Backpropagation | 975.67 | **1326.85** | 0.0003729 | **0.002018** |
| | Sub clustering | Hybrid | 985.54 | 1336.69 | 0.0003767 | 0.002027 |
| | | Backpropagation | 988.07 | 1336.99 | 0.0003774 | 0.002026 |

| Input | Clustering type | training methods | performance measures | | | |
|---|---|---|---|---|---|---|
| | | | RMSE | | RMSRE | |
| | | | Train | Test | Train | Test |
| | FCM clustering | Hybrid | 983.51 | **1330.65** | **0.0003762** | 0.002021 |
| | | Backpropagation | 997.39 | **1331.06** | **0.0003782** | 0.002011 |

TABLE III. ETH price forecasting result

| Input | Performance measures | | | |
|---|---|---|---|---|
| | RMSE | | RMSRE | |
| | Train | Test | Train | Test |
| ETH(k) | 76.29 | 112.92 | 0.0004061 | 0.002108 |
| ETH(k), ETH.D(k) | **76.24** | **112.89** | **0.0004055** | **0.002106** |

TABLE IV. Compare different algorithms for forecasting

| Predicted outputs | Prediction methods | Performance measures (RMSE) | |
|---|---|---|---|
| | | Train | Test |
| BTC(k+1) | ANN | 1365.07 | 1549 |
| | ANFIS | **997.39** | **1331.0** |
| ETH(k+1) | ANN | 101.3 | **91.08** |
| | ANFIS | **76.24** | 112.89 |

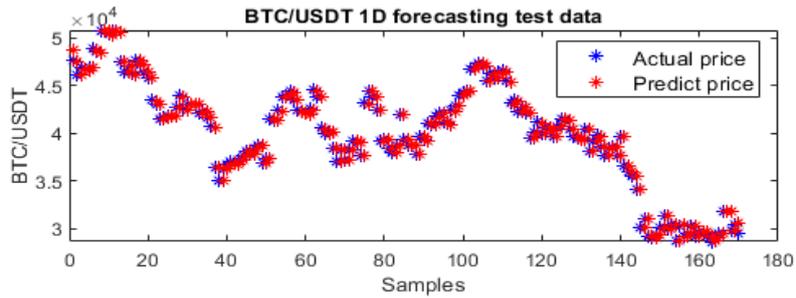

(a)

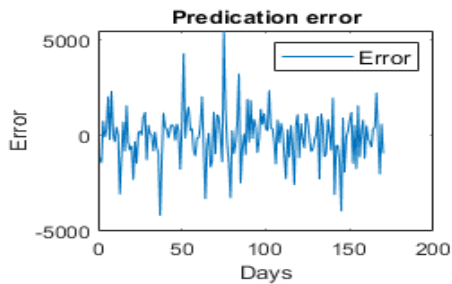

(b)

Fig.3. (a) Predicted price of BTC (red) compared to the real price (blue) in the ANFIS system in a daily time frame using two inputs of BTC and BTC.D in testing data; (b) prediction error.

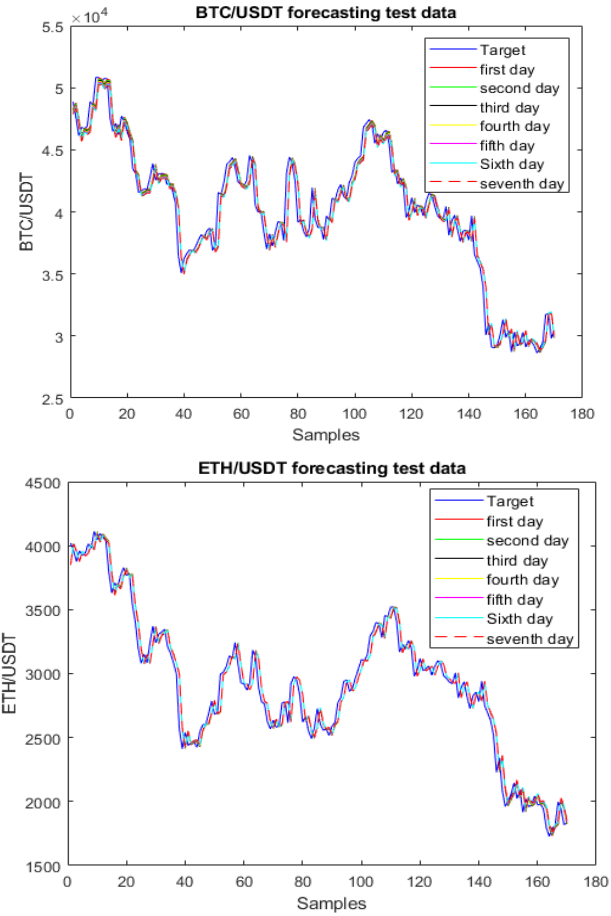

Fig. 4. Cryptocurrency price prediction until seven days, (a) predicted price of BTC until seven days compared to the real price (blue) in the ANFIS system in daily time frame using two inputs of BTC and BTC.D in testing data; (b) predicted price of ETH until seven days compared to the real price (blue) in the ANFIS system in daily time frame using two inputs of ETH and ETH.D in testing data.

## 5. CONCLUSION

This study shows that historical price data is a kind of time series data, so we can use time series methods to forecast prices. We use the ANFIS system, a time series method, to forecast the prices of cryptocurrencies like BTC and ETH in a daily timeframe for the following week. We first need to tune and assess the parameters of the ANFIS system to see which ones enhance the evaluation criteria for price forecasting. We split the BTC historical data into 90% training data and 10% test data and feed it to the ANFIS system. Based on Table 2, we see that the ANFIS system improves the evaluation criteria such as RMSE and RMSRE by using the FCM algorithm for data clustering and the backpropagation algorithm for data training. We also need to see if an indicator like BTC.D improves system performance. We test the system with two inputs, the price of BTC and the value of BTC.D, and based on Table 2, the evaluation criteria improve compared to using only our BTC historical data for forecasting. We do the same for ETH, and based on Table 3, the evaluation criteria improve when we use ETH.D along with the price of ETH. In Table 4, we contrast this method with the ANN method and see that the ANFIS method performs better in terms of RMSE for BTC prediction in both training and test data, but the ANN method does better for ETH prediction in test data. Finally, we use the ANFIS system, as shown in the block diagram in Fig. 2, to forecast the price for the next week. We use a system with several subsystems to train and learn all the dependencies such as the BTC.D index. This system forecasts the price in seven cycles for up to seven days, and the forecasted price in each cycle is input to the ANFIS system. The ANFIS system can forecast the price of the next day based on this price. Finally, based on Fig. 4, the forecasted price differs slightly from the actual price in each cycle.

## Authors

**Ali Mehrban** (Member, IEEE) received the B.S. degree from the IAUSR Tehran, Iran, in 2009 and the M.S degree from Newcastle University, Newcastle, UK, in 2011. From 2012 he has been IT & Network Consultant in private sector. He is currently a researcher in AI applications in communication systems and networking. His research interests include anomaly detection and resource allocation in networks using ML, cognitive radio networks, cooperative communication, edge computing, network security & 6G.